  \documentstyle[epsf]{article}
\def\bea{\begin{eqnarray}}
\def\eea{\end{eqnarray}}
\def\nn{\nonumber}

\def\beq{\begin{equation}}
\def\eeq{\end{equation}}
\def\ba{\beq\new\begin{array}{c}}
\def\ea{\end{array}\eeq}
\def\be{\ba}
\def\ee{\ea}

\def\Tr{{\rm Tr}}

\makeatletter
\newdimen\normalarrayskip              
\newdimen\minarrayskip                 
\normalarrayskip\baselineskip
\minarrayskip\jot
\newif\ifold             \oldtrue            \def\new{\oldfalse}
\def\arraymode{\ifold\relax\else\displaystyle\fi} 
\def\eqnumphantom{\phantom{(\theequation)}}     
\def\@arrayskip{\ifold\baselineskip\z@\lineskip\z@
     \else
     \baselineskip\minarrayskip\lineskip2\minarrayskip\fi}
\def\@arrayclassz{\ifcase \@lastchclass \@acolampacol \or
\@ampacol \or \or \or \@addamp \or
   \@acolampacol \or \@firstampfalse \@acol \fi
\edef\@preamble{\@preamble
  \ifcase \@chnum
     \hfil$\relax\arraymode\@sharp$\hfil
     \or $\relax\arraymode\@sharp$\hfil
     \or \hfil$\relax\arraymode\@sharp$\fi}}
\def\@array[#1]#2{\setbox\@arstrutbox=\hbox{\vrule
     height\arraystretch \ht\strutbox
     depth\arraystretch \dp\strutbox
     width\z@}\@mkpream{#2}\edef\@preamble{\halign
\noexpand\@halignto
\bgroup \tabskip\z@ \@arstrut \@preamble \tabskip\z@ \cr}%
\let\@startpbox\@@startpbox \let\@endpbox\@@endpbox
  \if #1t\vtop \else \if#1b\vbox \else \vcenter \fi\fi
  \bgroup \let\par\relax
  \let\@sharp##\let\protect\relax
  \@arrayskip\@preamble}
\def\eqnarray{\stepcounter{equation}%
              \let\@currentlabel=\theequation
              \global\@eqnswtrue
              \global\@eqcnt\z@
              \tabskip\@centering
              \let\\=\@eqncr
              $$%
 \halign to \displaywidth\bgroup
    \eqnumphantom\@eqnsel\hskip\@centering
    $\displaystyle \tabskip\z@ {##}$%
    \global\@eqcnt\@ne \hskip 2\arraycolsep
         $\displaystyle\arraymode{##}$\hfil
    \global\@eqcnt\tw@ \hskip 2\arraycolsep
         $\displaystyle\tabskip\z@{##}$\hfil
         \tabskip\@centering
    &{##}\tabskip\z@\cr}
\newfont{\hr}{msbm10}
\newfont{\ams}{msam10}
\begin{document}
\begin{titlepage}
\setcounter{footnote}0
\begin{center}
\hfill UUITP-7/95\\
\hfill FIAN/TD-19/95\\
\hfill ITEP/TH-9/95\\
\hfill {\it revised version}
\vspace{0.3in}

{\LARGE\bf  Towards effective topological gauge theories
on spectral curves}
\\[.4in]
{\Large A.Gorsky
\footnote{E-mail address: gorsky@vitep3.itep.ru}}\\
{\it ITEP, Moscow, 117 259, Russia}\\
\bigskip
{\Large A.Marshakov
\footnote{E-mail address:
mars@lpi.ac.ru}}\\
{\it Theory Department,  P. N. Lebedev Physics
Institute , Leninsky prospect, 53, Moscow,~117924, Russia\\
and\\
ITEP, Moscow 117259, Russia}
\end{center}
\bigskip
\bigskip

\begin{abstract}
We discuss a general approach to the nonperturbative treatment of quantum
field theories based on existence of effective gauge theory on auxiliary
''spectral" Riemann curve. We propose an effective formulation
for the exact solutions to some examples of $2d$ string models and
$4d$ supersymmetric Yang-Mills theories and consider their natural
generalizations.
\end{abstract}

\end{titlepage}

\newpage
\setcounter{footnote}0

Recently the development of matrix model technique
has lead to understanding the role of integrable structures in
string theory \cite{FKN,DVV,GMMMO,GKM}. One of the ways to
interpret this fact is to consider appearing integrable equations as
equations of motion in (hypothetic) string field theory and it looks natural
to search for a field-theoretical
formulation of this target-space theory directly, i.e.
without any referring to $2d$ world-sheet.
Moreover there exists a direct analog of this phenomenon for the $4d$
theory \cite{WS,WSGP} where four-dimensional space-time plays the role
of the world sheet and the exact solution can be given in similiar terms.
In both cases the most adequate formulation is given in terms of the
integrable systems of ''KP/Toda type".

In this letter we will argue that the appearence of integrable systems
can be treated within some unified framework and the nonperturbative
information can be read off from the topological effective gauge theories.
Usually
one starts with a bare theory defined on a ''world-sheet" and considers
Toda-type equations coming commonly from the
symmetries of a bare theory. Next, using the symmetries of the bare
model one can specify the solution
which corresponds directly to the nonperturbative regime and depends
on finite number of moduli parameters of the theory (at least we will
restrict ourselves for such cases). Finally, we will use the fact that
such solutions can be connected with topological gauge theories where
the auxiliary (from the world-sheet point of view) spectral curve
appears as a (part of) target-space. More strictly the spectral curve
of arising integrable system of Hitchin type \cite{Hi} is a
cover of a surface where the auxiliary gauge theory is defined. One may
hope that this is a generic scheme valid for a wide class of the
nonperturbative string and gauge field theories, below we will demonstrate
the main features of this scheme considering existing up to now $2d$ and
$4d$ examples.

First we will consider the simplest example when the role of the world-sheet
theory is played by a discrete matrix model and show that it can be
reinterpreted
in terms of a generalized $2d$ YM model. Then we will discuss a relation
between the (perturbed) $G/G$ gauged WZW model and some class of $2d$
topological theories.
Finally, we propose the effective theory for $4d$ $N=2$ SUSY YM and its
generalizations in terms of a so called holomorphic generalized YM model
on torus. The quantum characteristics of the effective theories like
spectrum and wave functions and their relevance to the nonperturbative
characteristics of bare models are also discussed.

\bigskip
{\bf 1.} Let us turn directly to the simplest example and consider a
{\it discrete} matrix model, where instead
of integration over the functions on world sheet $f(z,{\bar z})$ we are summing
over
matrices or functions $H_{ij}$ of two discrete variables.
The role of the reparametrizations of the
world-sheet is played by a sort of ''gauge" symmetry $H \rightarrow UHU^{-1}$
or in the infinitesimal form
$\delta H = [\epsilon, H]$
which is similiar to (complex) one-dimensional
world-sheet symmetries generated by $\delta _{\epsilon}H =
\partial _z^{|i-j|}H + ... $.
Since the gauge symmetry corresponds to reparameterizations the moduli space
of gauge connections is direct analog of moduli space of
complex structures in string theory.

The main observation of this part is that for discrete matrix model an
effective target-space
formulaton is given by a degenerate case of the generalized two-dimensional
YM theory. We should point out that in general the effective theory is defined
on a spectral curve though it is not seen here explicitely since we have a
degenerate case of a rational curve.  This is actually one of very important
questions and we will return to this discussion in detail below. Here the
role of a spectral curve is played by a circumference $0 < x < 2\pi $
where one can write the generalized
\footnote{which transforms to the
ordinary $2d$ YM action after integrating over $\Phi $ for $V(\Phi ) = \Phi
^2$. The stringy interpretation of the pure $2d$ YM theory was proposed in
\cite{Gross}.}
$2d$ YM action
\be\label{ymact} S = \Tr\int _{dxdt} \left(
F(A)\Phi - V(\Phi ) + A_i J_i \right) =
\nn \\
= \Tr\int _{dxdt} \left( \Phi
\partial _t A_x - V(\Phi ) + A_t(\partial _x \Phi + \left[ A_x,\Phi \right] -
I) \right)
\ee
giving the equations of motion
\beq\label{t} \partial _t \Phi +
\left[ A_t,\Phi \right] = J_x = 0 \eeq \beq\label{curv} F_{tx}(A) = \left[
\partial _t + A_t , \partial _x + A_x \right] = \partial _t A_x - \partial _x
A_t + \left[ A_t,A_x\right] = V'(\Phi)
\eeq
and the Gauss law
\beq\label{x}
\partial _x \Phi + \left[ A_x,\Phi \right] = J_t = I
\eeq
The Hamiltonian is
\begin{equation}\label{ham}
H = \Tr\int _{dx} \left(  V(\Phi ) + A_t(\partial _x \Phi + \left[
A_x,\Phi \right] -  I)\right)
\end{equation}
and it is possible to define additional (commuting) flows
\be\label{tk}
{\partial\Phi\over\partial t_k} = \{ H_k,\Phi \} \ \ \ \ \ \
H_k = \Tr \Phi ^k
\ee
where the Poisson bracket is induced by $\Omega _{YM} =
\Tr\int _{dx}\delta\Phi \wedge \delta A_x$.
In terms of integrable systems eq. (\ref{x}) is a reduction constraint, so
that $\Phi $ determined as a solution to (\ref{x}) becomes the Lax operator
of a discrete Calogero system.  Then (\ref{t}), (\ref{tk}) are the Lax
evolution equations in the corresponding finite-dimensional integrable
system. Finally, (\ref{curv}) is vacuum equation, or the initial condition
for the dynamical system (\ref{t}), (\ref{tk}).

Now the relation to discrete matrix model appears when one takes
$t$-independent solutions so that (\ref{t}) and (\ref{curv}) reduce to
\be\label{eqlim}
[ A_t , \Phi ] = 0
\nn \\
- \partial _x A_t + [A_t,A_x] = V'(\Phi ) = \sum kt_k \Phi ^{k-1}
\ee
The first one tells that $A_t$ is some function of $\Phi $
\footnote{actually $A_t$ is a
degeneration of the following expression in the elliptic case
$$
A_{t,ij}=\delta_{ij}\left(-\wp(\lambda ) + 2\sum_{k\neq i}\wp (x_i-x_k)\right)
+ 2(1-\delta_{ij})\Phi '(x_i-x_j;\lambda )
$$}
while the second says that the minima of the Hamiltonian (\ref{hm})
$ V'(\Phi ) = 0$ correspond to the solutions to
$[D_x, A_t] = [\partial _x - A_x, A_t] = 0$
so that $ A_t$ commutes with the ''shift" operator $ D_x$.
Thus, the first term in the action (\ref{ymact}),
vanishes and one gets the action on the vacuum solution
\beq\label{hm}
S_{vac} =  \Tr V(\Phi )
\eeq
Now one can easily notice that (\ref{hm}) looks similiar to the effective
action of discrete 1-matrix model \cite{GMMMMO}. To clarify the similiarity
let us after the redefinition of the variables
\be\label{todalim}
q_i = \phi _i + (i-1)\Delta  \ \ \ \
g = g_0 e^{\Delta \over 2}   \ \ \ \
\Delta \rightarrow \infty
\ee
take the limit \cite{Ino} from
Calogero to nonperiodic Toda chain
\be\label{ino} H
= \sum {1\over 2} p_j^2 + g^2 \sum _{i>j} \wp (q_i - q_j) \rightarrow
\nn \\
\rightarrow
\sum {1\over 2} p_j^2 + g^2 \sum _{i>j} \sinh ^{-2} (q_i  - q_j)
+ const
\rightarrow  \sum {1\over 2} p_j^2 + g_0^2 \sum
e^{\phi _j - \phi  _{j+1}}
\ee
Keeping $g_0 \rightarrow\infty $ means that
the Toda limit with $g^2 = \nu (\nu -1)$ implies $\nu \rightarrow \infty $.
The Toda chain $\tau $-function
\beq
{\tau _N\over \tau _0}=
\det C _{N\times N} =
e^{\sum _{i=1}^N\phi _i} = {\tau _N
(t)\over\tau _0 (t)} \sim e^{\Tr\int a_x dx} \sim \det \exp {e^{\int a_x dx
}}
\eeq
is the determinant of the moment matrix (see for example \cite{KMMOZ}),
\cite{KMMM-T}
\footnote{being for the 1-matrix model
$$
C_{ij} = \int d\lambda e^{- V(\lambda )} \lambda ^{i+j-2}
$$}
which in this example is the monodromy matrix
of the YM field $ a_x = diag(\phi _1 ...\phi _N)$.
Now, string equation
\be\label{wi}
 \sum
_{k>0} kt_j {\partial\over \partial t_{k-1}}\tau _N (t) = \sum kt_k
{\cal H}^{k-1} =
\sum _{k>0} kt_k \Tr L^{k-1} = 0
\ee
can be interpreted as an equation for the Lax operator
$L[\phi (t)]$
and the matrix model {\it effective} action \cite{GMMMMO}
\beq\label{mamoea}
S =   \Tr\sum _k t_{k} L^{k} + S_0 + {1\over 2}\sum _n \phi _n
= S_{vac} + {1\over 2}\Tr A_x
\eeq
almost coincides with (\ref{hm}) after Toda reduction. The additional
term $\Tr A_x$
\footnote{Coming from
$$
\sum _{k>0} kt_k \langle n|\lambda^{k-1}|n\rangle =
\sum _{k>0} kt_k \left(L^{k-1}\right)_{nn} = 0
$$
and
$$
\left({\partial V\over\partial L}\right)_{n-1,n} = \sum _{k>0} kt_k
\left(L^{k-1}\right)_{n-1,n} = {n\over L_{n,n-1}} =
n e^{-{1\over 2}(\phi _n - \phi _{n-1})}
$$}
is related to more tiny effects which will be discussed later.
This actually demonstrates the necessity of quantization of the effective
theory.

Consider now the solutions to string equation in the simplest
cases.  From (\ref{wi}) it follows for $ t_k = 0$ for $ k\geq 2$
we have
\begin{equation}\label{momentum}
\sum_{i}t_2{\partial {\phi _i}\over \partial t_1} \sim \sum{p_i} =
Nt_1
\end{equation}
having the sense of total momentum ''flow" in $t_1$-direction for zero higher
times
\footnote{
For generic point in the space of times (or coupling constants) one
has instead of (\ref{dse2}) the condition
$$
\sum{kt_kH^{k-1}} = t_1N + 2t_2P + 3t_3E + ... = 0
$$
which for example for nonzero $t_3$ and zero higher times has a sense
of ''energy" dissipation.}.
String equation results in a peculiar  $\Delta $-dependence for
the YM coupling constant. Using relation \cite{GMMMO}
\beq\label{dse2}
n = \sum _{k>0} kt_k {\langle n-1 |\lambda ^{k-1} |n \rangle\over
\langle n|n \rangle}
e^{\phi _n - \phi _{n-1}}
\eeq
giving for only $t_2 \neq 0$
\be\label{move}
e^{\phi _n - \phi _{n-1}} = {n\over 2t_2}
\nn \\
\phi _n = \log n! - n\log 2t_2
\ee
one can single out the linear piece and compare with
(\ref{todalim}). The result gives
\beq\label{deltadep}
t_2 = g^2(\Delta ) = e^{\Delta }
\eeq
Now we will discuss another close example concerning the effective
description in terms of the gauged $G/G$
\footnote{For example $G = SU(N)_k$ or $U(N)_{k-N,k} \equiv SU(N)_{k-N}
\otimes U(1)_k$}
Wess-Zumino-Witten model which is a direct generalization of $2d$ YM
system \cite{gn1}.
The Lagrangian defined on a (''spectral") surface has the following
form
\be\label{action}
S = k S_{WZW} (g) + \frac {ik}{2\pi }\Tr\int ( A_z g^{-1}{\bar\partial}
g + g\partial
g^{-1} A_{\bar z } +gA_z g^{-1} A_{\bar z } - A_z A_{\bar z}  )
\nn \\
+ V(g) + \nu \omega _{\bar z}A_tJ_z
\ee
with the Hamiltonian
\begin{equation}\label{relham}
V(g) = {1 \over 2}\sum ( t_{k}\Tr g^{k} + {\bar t}_k \Tr g^{-k})
\end{equation}
The equations of motion are
\be\label{em}
g^{-1}{\bar\partial} g + g^{-1}A_{\bar z}g -A_{\bar z} = J_{\bar z}
\nn \\
F_{z,\bar z }\left[A_z , g^{-1}\partial g + g^{-1}A_{\bar z}g \right]
\equiv \bar\partial A_z - \partial (g^{-1}\bar\partial g + g^{-1}A_z g)
\nn \\
+ \left[ A_z, (g^{-1}\bar\partial g + g^{-1}A_z g) \right) = g^{-1}V'(g)
\ee
and generalized Gauss law is
\begin{equation}\label{relgl}
g\partial g^{-1} + gA_z g^{-1} - A_z = J_z
\end{equation}
The observables in the theory are $\Tr_V g$ or the Wilson loops in
different representations.

Now, let us remind how the Ruijsenaars dynamics appears in the $G/G$
theory \cite {gn1}.
Again, one should consider the special sources $J_{ij}=1_{ij}-\delta_{ij}$
and after the resolution of (\ref{relgl}) it appears
that in the gauge $A=diag(q_1,\dots ,q_n)$ $g$ becomes
the Lax operator for the trigonometric (or hyperbolic) Ruijsenaars system
\beq\label{triru}
L_{ik}=e^{p_i}\prod_{i\neq j}
\frac{\sin [(\frac{\pi}{k})(q_i-q_j+\frac{1}{k}]}
{\sin [(\frac{\pi}{k})(q_i-q_j]}
\eeq
where $q_i$ and $p_i$ are
canonically conjugated variables.

On its own the Ruijsenaars system is related with the pole solution to the
2D Toda lattice \cite{kriza}. Consider the elliptic
solutions of the Toda lattice in (discrete) $x$ variable,
namely let us assume that $\phi(x+\eta,t,\bar{t})-\phi(x,t,\bar{t})$
is an elliptic function of the variable $x$. Then $\phi$
is a solution of the Toda equations if and only if the poles
of the solutions $x_i$
\beq\label{fiel}
e^{\phi} =\prod_{i=1}^N\frac{\sigma(x-x_i+\eta)}{\sigma(x-x_i)}
\eeq
move  with respect to the linear combination of the time variables $t$,
${\bar t}$
according to the Hamiltonian equations of motion with
$H= \Tr L + \Tr L^{-1}$
where $L$ is the Lax operator for the elliptic Ruijsenaars
system. Here we are interested in the trigonometric degeneration
of this relation. Note that comparing (\ref{triru}) with (\ref{fiel}) we
see that $\eta = \frac{1}{k}$.

Now one can remind that \cite{WIT} that GWZW can be considered as a low energy
limit of the $2d$ SUSY topological $\sigma$-model with $U(N)$ gauge
symmetry and $kN$ chiral matter multiplets, so that if one integrates out
the chiral matter the resulting low energy effective action is given by
GWZW theory.

Let us turn now to the integrable structure behind the topological
$\sigma$ models, where \cite{CeVa} the underlying
integrable system is again 2D Toda lattice.
The main input for the integrable structure (see for example \cite{Dubrovin}
and references therein) is the
(perturbed) quantum cohomology ring generated for $CP(N)$ by a
factorization over
\beq
x^n = \beta
\eeq
where $\beta$ is the action calculated on an instanton.
The corresponding coupling
constants play the role of the "times" in the integrable
hierarchy, they contain the coupling to the
K\"ahler class $\log\beta = \frac{i}{g^2} + \theta$ where
$g$ is the bare
coupling constant of the $\sigma$ model and $\theta $ is bare
$\theta$-term.

The Toda lattice variables
\beq\label{shift}
\phi_{i}=lng_{i\bar{j}}-\frac{2i-N+1}{2N} ln|\beta|^{2}
\eeq
are the eigenvalues of the ground state
metric \cite{CeVa} related with the nonabelian Berry connection and can be
extracted from the two-point
topological correlators. The relation (\ref{shift}) reminds a lot (\ref{move})
from the previous example.

It is necessary to mention that unlike \cite{CeVa},
where reduction to the radial coordinate $|\beta|$ with the Painleve-3
solution arose due to the equal number of instantons and
anti-instantons, we do not impose such a condition and obtain another
solution to the Toda lattice equation. If one tends the level of the theory
to infinity there will be the topological YM theory on one side and
the continious KP equation on the other.

\bigskip
{\bf 2.} It is known that nonlinear $\sigma$ models play the role of
the playground for the nonperturbative approaches
to $4d$ QCD. Recently \cite {WS} the nonperturbative
solution to $N=2$ SYM theory has been derived
which provides exact mass spectrum and the low-energy effective action.
It was shown in \cite{WSGP}
\footnote{and this idea was developed in \cite{Tak}, \cite{MartWa},
\cite{WiDo}, \cite{EgYa}}
that there exists an integrable
structure behind this solution so that the massive spectrum is determined
by the periods and effective action is a logariphm of the $\tau$-function
of the Whitham hierarchy.

The main
step of the solution is the determination of the spectral curve
with some additional structure which carries all nonperturbative
information about the theory. The spectral
curve plays the role of the initial data (boundary condition)
and we are going to define an effective topological theory on this curve.
Having some picture for the $2d$ theory one can think that
in $4d$ theory
the variables $t$, $\bar{t}$, $x$ are also the same combinations
like $e^{\frac{i}{g^{2}}+\theta}$, $e^{\frac{i}{g^{2}}-\theta}$
and the Chern number, but the accurate analysis needs of course the
determination of the topological subsector. We will not discuss this
important point here, however as it was mentioned in \cite{WSGP} the desired
$SU(N)$ curve found in \cite{argires} arises in a periodic Toda-chain system.

In what follows we are going to consider the periodic Toda chain as a
degenerate case of a certain Hitchin-type \cite{Hi} dynamical system on
target-space torus and related with the affine $sl(N,{\bf C})$ algebra.
We will try to demonstrate that these sort of formal generalizations are
also quite interesting from physical point of view.

Let us start with the {\it
holomorphic } YM theory action define on a torus
\beq\label{ymact1}
S = \Tr \int _{d^2zdt} \omega _{\bar{z}} F_{tz}(A)\Phi - \omega _z V(\Phi )
+ \nu A_t J_{\bar z}\omega _{\bar{z}}
\eeq
where $\omega _z$ is the holomorphic 1-differential,
and equations of motion generalize (\ref{t}), (\ref{curv}), (\ref{x})
and (\ref{em})
\be\label{t1}
\partial _t \Phi + \left[ A_t,\Phi \right] = 0
\nn \\
F_{tz}(A) = \left[ \partial _t + A_t , \partial _z + A_z \right] =
\partial _t A_z - \partial _z A_t + \left[ A_t,A_z\right] = V'(\Phi)
\ee
\beq\label{x1}
\partial _z \Phi + \left[ A_z,\Phi \right] = J_z
\eeq
The Hamiltonian is now $H = \Tr\int d^2z Q(\Phi)$
where $Q(\Phi)$ is an invariant
polynomial on the Lie algebra $sl(N,C)$, and finite-dimensional
dynamical system appears as before after the resolution of the Gauss law
constraint so that \cite{gn2} $\Phi$
occures to be a Lax operator of an elliptic Calogero-Moser system or a
particular case of the Hitchin system on torus. Now
the key point is that the periodic Toda chain arises as a limit
of the elliptic
Calogero system and hence of the corresponding effective topological
field theory on target space.

Starting from the initial elliptic Calogero
(\ref{todalim}) we have to proceed now in a slightly more delicate way than
before. One has
to introduce the new Toda variables by $q_i = \phi _i + (i-1)\tau$
and put \cite{Ino}
\be\label{pertodlim}
g^2 = e^{\tau} = e^{{\tilde\tau}\over N}
\ \ \ \
{\tilde\tau} \rightarrow \infty
\ee
As a result we obtain the periodic Toda chain when the Lax operator of
elliptic Calogero
\beq\label{lelcal}
L_{Cal,ik}(z) = \delta_{ik}p_{k}+i(1-\delta_{ik})
\frac{\sigma(q_{i}-q_{k}+z)}{\sigma(z)\sigma(q_{i}-q_{k})}
\eeq
reduced to
\be
L_{Toda,jk}(w) = \delta_{jk}p_k + \delta_{j,k+1}\exp(q_j-q_{j-1})-
(i)^N w\delta_{jN}\delta_{k1}
\nn \\
-i^{-N}w^{-1}
\delta_{j1}\delta_{kN}\exp(q_1-q_N)
\ee
The Calogero spectral curve which is in this case
the $N$-sheet covering of the torus defined by
\begin{equation}\label{ntorus}
\det _{ij} \left( L_{Cal} (z) - \lambda \right) = 0
\end{equation}
degenerates to
\be\label{todacurve}
w + {1 \over w} = 2P_{Toda}(\lambda )
\ \ \ \
y^2 = P_{Toda}^2 - 1
\ee
where $2y = w - w^{-1}$.
Note that
if we keep the coupling constant in Toda after the short calculation
one gets $\Lambda^N$ instead of unity in the last term in
(\ref{todacurve}). This clearly indicates that this term plays
the role of the instanton correction just as in $2d$ case.

Thus it turns out that the Seiberg-Witten (and their $SU(N)$
generalization) curves are special limits of the $N$-sheet
covering of the torus. Moreover there exists a certain generalization
\footnote{related hopefully to the quantum affine algebras}
again having the form of (\ref{ntorus})
where $L_{Cal} (z)$ now should be replaced by the Lax matrix of the
Ruijsenaars system
\be\label{Rlax}
L_{ellR-s,jk}(z) = \exp(\beta
p_j)V_j(q)\frac{\sigma(q_j-q_k+ z)\sigma(\eta )}{\sigma(z )\sigma
(q_j-q_k+\eta)}
\ \ \ \
V_j = \prod_{j\neq k} f(q_j-q_k)
\nn \\
f^{2}_{ellR-s}(x) = \sigma^2(\eta)\left(\wp (\eta)-\wp (x)\right)
\ \ \ \
f^{2}_{rel-Toda}(x) =1+\tau^{2}\beta^{2} e^x
\ee
where the last expression corresponds to the degenerate periodic relativistic
Toda case \cite {rs}.
We conjecture that generic Calogero -
Ruijsenaars type curve might correspond to some $4d$ gauge systems
coupled to special matter. The case with the fixed flavour number and
arbitrary mass scale of matter, being a special case of (\ref{lelcal}) is
discussed in \cite{WiDo}, and one might expect the appearence of more general
curves of the (\ref{ntorus}) type for generic $N_f$.
Typically the level of Kac-Moody algebra is related with the number of
the fermionic flavours so we can conjecture that such
curves appear if one adds the $k = N_f$ flavour matter to the theory on
the world-sheet. It would be very important to get the proper world sheet
theory because it would provide some view on a role of the quantum
group and Sklyanin algebra in essentially $4d$ theories.

The generalization from Toda to Calogero case can be also thought of as
follows. One can
think that the papameter $\tau$ serves as a measure of the "vacuum
transition amplitudes" and the reduction to Toda system means that the
amplitude is
''small" and only the transitions between the neighbour vacua survive.
The transition from Calogero to Toda looks like the transition from the
''liquid" phase to the ''crystal" one.

Note that there are other possible generalizations of the spectral curve.
One can add the internal degrees
of freedom to the ''particles" -- that would lead to the ''nonabelian"
integrable systems similiar to those of \cite{kriza}, consider  theory on
higher genus
curves (see for example \cite{nerso}) or add additional marked
points.  We conjecture that all these possibilities can be realized if
one starts with more complicated bare theory, i.e. takes more initial coupling
constants or consider more general symmetry breaking scheme.

\bigskip
{\bf 3.}Finally, we will try to discuss the meaning of quantization of arising
effective theory. First let us mention that in $2d$ case from string theory
point of view this is a rather natural procedure which should lead to the
second-quantized string field theory in its original sense. In the $4d$
situation this is not as clear but we would still argue that the quantum
characteristics of the effective topological gauge theory (wave functions,
spectrum etc) are relevant for the description of bare model.

Let us start again from the most simple example of discrete matrix model.
As we have tried to argue above the effective theory can be represented in
the form of generalized YM model which reduces on particular solutions to
the effective discrete matrix model action written in terms of the Lax
operator. Let us notice first that the wave function of the generalized
YM theory in the limit $\nu\rightarrow 0$ acquires the form of the GKM
partition function \cite{gn1,GKM2} giving for this case the generating
function for the exact correlators in $2d$ topological gravity models.
This analogy as well as relation of the Ruijsenaars wave functions (or
the Harish-Chandra functions) to the soliton S-matrices \cite{BaBe} allows
one to hope that the wave functions of the systems considered above can
be treated as generating functions for nonperturbative correlators.
These wave
functions are known to be certain one-point conformal
blocks on torus \cite{Etingof},
\be\label{etin}
\Psi _{\lambda}({\bf t},q) = {\Tr _{\lambda}\left( Vq^{L_0}e^{\bf tH}
\right)\over \Tr_{-\rho}\left( V q^{L_0}e^{\bf tH}\right)}
\ee
being the solutions to the
Knizhnik-Zamolodchikov-Bernard equations in the $\widehat{sl(N,{\bf C})}$
WZW theory at critical level $K =-C_V$.  The limits (\ref{todalim}),
(\ref{pertodlim}) imply that one has to put $C_{\lambda} = e^{\Delta}$ for
$\Delta \rightarrow\infty $ which means that for the Toda theory we end up
with the infinite-dimensional representation. In such limit the wave
function (\ref{etin}) turns into the periodic Toda-chain wave function
which should be related with the generating function for the
nonperturbative $4d$ theory.

Another ($\nu\rightarrow\infty $) limit giving rise to the discrete matrix
model itself can be reformulated in terms of the following Lax operator
$\Phi (A_x)  \rightarrow L(\phi ) \rightarrow W(X) = X + {n\over X}$
playing the role of LG superpotential in GKM-type theories. In this
formulation the corresponding action \cite{KM} $V(X) = \int dX W(X) =
{1\over 2}X^2 + n\log X $
naturally separates into quadratic classical ''YM" piece and the logariphmic
one-loop correction and serves as a potential in the exact partition function
\cite{KMMM-T}. Similiar superpotentials arise in the case of topological
$\sigma$-models \cite{CeVa}.

One could consider here two naively different types of quantization: the
first
deals with the Whitham type deformation of the spectral curve \cite{Kri},
another quantizes the gauge theory on the undeformed surface. The first one
can be identified with the renormalization group flow \cite{WSGP} in world
sheet theory and in $2d$ case leads to the quantization of finite-gap
potentials \cite{NKM} and to the GKM-type theory.
From the point of view of gauge theory it leads to the appearence of
Grassmannian-like fermionic constructions when effective action
arises as a result of the integration over fermionic degrees of freedom
on spectral curve.

The second
way should have a meaning of the derivation of a
kind of "effective action" for world sheet theory before renormalization.
For example in the $SL(2)$ case the equations of
motion
\be
{d^2\phi\over dt^2} + g\sin\phi = 0
\ \ \ \
H = {1\over 2}\left({d\phi\over dt}\right)^2 - g\cos\phi
\ee
can be easily integrated giving
\be
t = {1\over\sqrt{2g}}\int {dz\over\sqrt{(z-u)(z^2 - 1)}}
\ee
with $u \equiv {H\over g}$ thus giving rise to the $N=2$ SUSY YM spectral
curve \cite{WS,WSGP}. This theory should be considered as an effective one
coming from integration over the (Grassmannian) fermions so that the full
spectrum can be considered as a sum of the ''fermionic one" where the
fermions are solutions of auxiliary linear problem and the perturbations
of the above classical solution. The first part gives rise to the
Seiberg-Witten terms of the type $E_{ferm} \sim \int dS = \int Edp$ and one
can hope that the spectrum of auxiliary theory itself might
be related with the spectrum of ''extra states" of original theory similiar
to pomerons.The whole spectrum of exitations will be considered elswhere.

Let us finally discuss the quantization from more general point of view.
As usual
we deal with the world sheet theory which after integration over the
world-sheet variables gives us some effective target-space model where
we interest in the low-frequency ''visible" part of the spectrum. The
string equation gives us a renormalization group equation while
a particular critical point corresponds to some $\beta $-function fixed
point condition. Now, at any concrete fixed point we end up with the
{\it finite-dimensional} subspace in the space of the coupling
constants, i.e. in other words each fixed point correspond to a
renormalizable theory in the common field-theoretical language (one
should not add any additional counter-terms to renormalize the theory).
The string equation considered in the vicinity of such critical point
acquires the form of the characteristic solution to particular Whitham
equations. The main subtle point here is that the finite-dimensional
Whitham integrable system is singled out by some natural requirement of
periodicity and it is natural to ask what is the sense of the
periodicity condition in field theory. The only possible candidate one
has in mind is the $\theta$-parameter (or its conjugate $K$).

\bigskip
In this letter we have proposed the approach when the integrable structure
corresponding to the nonperturbative solutions of some $2d$ and/or $4d$ world -
sheet theory is encoded in the effective topological gauge theory on some
''spectral" curve. This effective topological theory governs the dynamics
on the moduli space of bare model which together with the spectral curve
describes the solution. We have conjectured that the quantization of the
effective theory is not meaningless from the point of view of the original
problem hence the reformulation of the problem in canonical field-theoretical
language is attractive from many different points of view.

The reason for the appearence of an effective gauge theory can be commented
in the following way. If there exists a degenerate vacuum state in the bare
theory one can define a (non Abelian) Berry connection. It is this connection
which can be considered as a connection on ''spectral curve" discussed above.
The presence of marked points on spectral curve (= ''dispersion law") is
related to the level crossing and corresponds to the vanishing of mass gap
for quasiparticles providing some symmetry breaking mechanism in bare
theory.

Of course, our understanding of many points discussed above is far from being
complete and we are going to return to these problems elsewhere.

We are indebted to I.Krichever, A.Mironov, A.Morozov, N.Nekrasov and V.Rubtsov
for useful discussions. We are grateful to J.Ambjorn for the hospitality in
the Niels Bohr Institute were this work was initiated and to
A.Niemi for
the hospitality in the Uppsala University where the
main part of this work was done. This work was partually supported by ISF
grant MGK000 and by the grants of Russian fund of fundamental research
RFFI 93-02-03379 and RFFI 93-02-14365. The research of A.M.
was also supported by NFR-grant No F-GF 06821-305 of the
Swedish Natural Science Research Council.

\bigskip
{\bf Note added.} When this paper was finished we have learnt about
\cite{Mart} where related topics are discussed.

\end{document}